# Massive Open and Online Courses (MOOCs) and Open Education Resources (OER) in Singapore

Victor Lim, Lawrence Wee, Shannalyn NG and Jessica Teo

Educational Technology Division, Ministry of Education, Singapore

## ABSTRACT

This paper looks at the increasing popularity of Massive Open and Online Courses (MOOCs) and Open Education Resources (OER) offered in Singapore. Despite being a relatively new phenomenon, the Singapore government has collaborated with different organisations to improve the quality and accessibility of MOOCs, and many Institutions of Higher Learning (IHLs) are spearheading efforts to improve OER to facilitate greater public access to educational resources. It will also explore the benefits and the potential problems that MOOCs and OER face. For example, both MOOCs and OER are able to lower the costs of university-level education and increase public access to such courses. They also provide skills and job training for members of the public, and encourage life-long learning. However, both MOOCs and OER may not be sustainable in the long run, as the financial gains of both may not be able to cover the cost of mounting them. Each system also has its own set of problems. For example, there is currently a lack of formal structures to guarantee the quality of MOOCs offered. MOOCs also tend to have low completion rates and there have been issues regarding plagiarism with the use of MOOCs as learning platforms. OER challenge traditional copyright policies, while the lack of sustainable funding prevent it from being adopted more widely. Even though both systems may potentially transform the traditional education system, a deeper understanding of MOOCs and OER, as well as their implications on learning will be useful.

**INTRODUCTION**

Recent developments in technology, the global ubiquity of devices, and the increase of Internet users worldwide have ushered in new educational phenomena in the form of Massive Open and Online Courses (MOOCs) and Open Education Resources (OER). They promise learners abundant, cheaper and accessible opportunities to education with the existence of open educational resources and tools in the virtual world (Kim, 2015). They also offer innovative approaches to the development, dissemination, and utilisation of knowledge in teaching, learning and research. This article highlights some developments and applications of MOOCs and OER in Singapore, and discusses the promises and challenges that these new educational phenomena present.

**MASSIVE OPEN ONLINE COURSES (MOOCS)**

MOOCs, a recent development in distance education, were first introduced in 2008 and emerged as a popular mode of learning in 2012 (Kim, 2015). The development of MOOCs is rooted in the ideals of openness in education; that knowledge should be shared freely and the desire to learn should be met without demographic, economic, and geographical constraints.

MOOCs' potential of 24 hour access to information, self-paced learning, and cost effectiveness have attracted millions of learners across the world. MOOCs can reach out to a massive group of participants online and allow for interaction among diverse learners across ages, cultures, and nationalities. As a result, MOOCs have received much attention from the media and have gained significant interest from institutions of higher learning (IHLs). There are now more than 4200 MOOCs offered by more than 500 universities (Valenzuela, 2016).

In contrast to traditional university online courses, MOOCs differ in two key features[1]:

- Open access - anyone can participate in the online course for free
- Scalability - courses are designed for a massive number of participants

However, MOOC providers interpret these features in different ways and to varying degrees. Some MOOCs are massive but not open; some are open but not massive. There are also issues with regard to the licensing and permissions of current MOOC provision and how these are aligned to the OER community.

MOOCs are not only extensions of existing online learning approaches, they also offer an opportunity to think afresh about new business models that include elements of open education (Li & Powell, 2013). MOOCs can potentially drive down the cost of university-level education and radically disrupt existing models of higher education. As a result, there is a growing interest and significant enthusiasm of MOOCs from Governments, institutions and business associations. A growing number of institutions have been involved in engaging and experimenting with MOOCs with the end goal of expanding access with greater potential of showcasing and marketing MOOCs to grow new income streams. Many IHLs around the world have responded, in varying degrees, to MOOCs. IHLs have come together to make their courses available online by setting up open learning platforms, such as Coursera, and edX. These platforms have been launched in collaboration with Ivy League universities which offer online courses for free and with a small fee for certification. Of all the providers of MOOCs around the world, Coursera leads in enrolment (35%) followed by edX (18%) (Valenzuela, 2016). Multinational corporations, such as Pearson and Google, are also planning to adopt a MOOC-based approach as a part of their foray into the higher education sector (Valenzuela, 2016).

---

[1] https://en.wikipedia.org/wiki/Massive_open_online_course

MOOCs can generally be categorised into two distinct types - cMOOC and xMOOC. cMOOC emphasises the development of understanding and knowledge through forum discussions and collaboration on joint projects, guided by connectivist theory. xMOOC resembles teacher-centric lectures, which are guided by behaviourist theory.

Motivations for learners to participate in MOOCs are varied. While the participation rate of MOOCs remains high, the completion rate is low. The market value of certification, short of a credit as part of traditional institutional awards, has yet to be determined.

**MOOCS IN ASIA**

There have been some instances where countries have encouraged the development and use of MOOCs within education systems in Asia. While MOOCs have only been popularised fairly recently (in 2012), there are now over 70 universities offering MOOCs to adult learners in Asia. The more popular providers are Coursera, Udacity and edX. It has been reported that 30% of the Asian population has since registered for a MOOC (Valenzuela, 2016). Most Asian MOOC users see these courses as a way to help them gain specific job skills, prepare for future work, and as part of professional certification. Asian users who have completed a MOOC are generally educated learners with professional degree.

**Singapore**

MOOCs play an important role in the next phase of education and skills development in Singapore. As part of encouraging Singaporeans to develop deep skills through life-long learning, the Singapore Government has implemented SkillsFuture, a nation-wide movement, in support of the Continuing Education and Training (CET) Masterplan[2]. Through SkillsFuture, Singaporeans, who are over the age of 25, are given SkillsFuture credits worth SGD500 to use for any accredited training programmes. Mid-career enhancement for citizens over 40 years old would have 90% of their course fees subsidized (Valenzuela, 2016). These SkillsFuture credits can be used for selected MOOCs on platforms such as Coursera, Udemy, and SIM University. It has been reported that MOOCs form 12% of all SkillsFuture Credit-eligible courses[3]. About 6% of Singaporeans have utilised their SkillsFuture Credits on MOOCs, and the majority of whom are in the 25 to 39 age group[4].

In addition, the Singapore Workforce Development Agency (WDA) and the Institute of Adult Learning (IAL) work closely with adult educators, business leaders, human resource developers and policy makers to transform the Continuing Education and Training (CET) sector. In recent years, IAL has partnered Canvas and Udemy to create and deliver MOOC offerings[5]. The European Union Centre in Singapore also offers EU courses through MOOCs (Cheah, 2016). Since 2014, universities in Singapore, such as the National University of Singapore, and Nanyang Technological University have offered MOOCs on platforms such as Coursera. The credits gained can be used as part of a student's qualification for a degree[6].

---

[2] http://www.mom.gov.sg/employment-practices/skills-training-and-development/refreshed-cet-masterplan
[3] http://www.ssg-wsg.gov.sg/new-and-announcements/08_Jan_2017.html
[4] Ibid

## ISSUES AND CHALLENGES FOR MOOCS

MOOCs present many opportunities to disrupt traditional higher education modes of learning and facilitate lifelong learning for adults. However, there are issues and challenges to overcome such as the quality of courses and completion rates, as well as the award and recognition of credit, pedagogy, and sustainability[7].

### i. Quality and Completion Rates

Critics have noted that MOOCs cater to a select group of participants who are already interested and motivated to learn via online platforms. MOOCs also demand a certain level of digital literacy from participants. These have led to concerns of inclusivity and equality of access (Li & Powell, 2013).

Typically, there is a lack of formal quality assurance for MOOCs. Crowd-sourcing seems to be the preferred way to ascertain quality. Courses are often evaluated by the participants, resulting in league tables that rank the courses by the perceived quality of the offering (Li & Powell, 2013). As such, courses that rate poorly will either disappear due to lack of demand, or survive by improving course quality in response to poor ratings. Other ways courses are evaluated include informal feedback comments from participants on social media.

The low rates of completion for MOOCs have been a point of controversy as well. Meyer (2012) reported that the dropout rates of MOOCs offered by Stanford, MIT and UC Berkley were 80-95%. For example, only 7% of the 50,000 students who took the Coursera-UC Berkeley course in Software Engineering completed the course. There is a similar reported dropout rate in Coursera's Social Network Analysis class, where only 2% of participants earned a basic certificate and 0.17% earned the higher level programming with distinction certificate. However, some have argued that whether or not these rates matter depends on the purpose of the MOOCs in the first place. If the purpose was to provide access to free and high-quality courses from elite universities and professors to a wide range of learners, then high dropout rates may not be a primary concern (Li & Powell, 2013). However, it would still be useful to improve the retention rates of MOOCs. Studies have been conducted to find out why and at what stage students drop out of courses.

---

[5] https://www.udemy.com/user/ialsg/
[6] http://www.channelnewsasia.com/news/singapore/after-successful-run-of/2267868.html
[7] http://publications.cetis.org.uk/wp-content/uploads/2013/03/MOOCs-and-Open-Education.pdf

### ii. Award and Recognition of Credit

Most MOOCs use quizzes as their main instrument of assessment – short multiple choice questions with automated answers for feedback. Some may offer other types of assessment that require open responses. However, due to limited resources, it is not realistic for thousands of essays to be graded by a lecturer. As a result, some MOOCs rely heavily on peer engagement and assessment to support the individual student's learning process. Coursera, for example, includes the submission of essay style answers which is graded through peer assessment.

There have been concerns about cheating and plagiarism with online learning, particularly when courses are eligible for academic credits. Measures taken to avoid this include sit-down written examinations as part of a MOOC. For example, Coursera worked with Pearson test centres to provide in-person examinations[8].

### iii. Pedagogy

Most available MOOCs are in the xMOOCs format. However, xMOOCs have been criticised for adopting a knowledge transmission model of learning. They are considered to be technology-enriched traditional teacher-centred instruction[9]. Coursera gives the institution the freedom to design their courses within broad guidelines. However, the institution may not have the necessary resources and manpower to design quality xMOOCs, as they are more time-consuming to produce and require much more planning and coordination.

cMOOCs, on the other hand, provide great opportunities for non-traditional forms of teaching approaches and learner-centred pedagogy where students learn from one another. Online communities 'crowd-source' answers to problems, and create networks for learning in ways that seldom occur in traditional classrooms in universities. For example, institutions like MIT and Edinburgh University are using MOOCs as an experimental venture to participate in emerging pedagogical models, exploiting peer support and using peer assessment techniques[10].

### iv. Sustainability

Organisations are still developing approaches to generate sustainable profit from MOOCs. Some common approaches to generate revenue include: charging students a fee for certificates of participation, completion or transcripts; providing premium services such as recruiting tools that link employers with students who have shown ability in a given area; and philanthropic donations from individuals and companies.

However, it remains a challenge for partner universities to generate income in these ways. In established business models, universities have control over the customer value proposition in that they provide any recognition of learning and set tuition fees (Li & Powell, 2013). For MOOCs, most participating institutions may choose not to offer credits as part of traditional awards for these courses. This may be due to concerns about the quality of the courses and the downside risks posed to their branding. Some may also resist charging fees as doing so contradicts the ideals of MOOCs which is to offer free education to all.

---

[8] http://publications.cetis.org.uk/wp-content/uploads/2013/03/MOOCs-and-Open-Education.pdf
[9] Ibid
[10] http://publications.cetis.org.uk/wp-content/uploads/2013/03/MOOCs-and-Open-Education.pdf

## OPEN EDUCATIONAL RESOURCES

Open Educational Resources (OER) are teaching, learning, and research resources that reside in the public domain or have been released under an intellectual property license that permits their free use and re-purposing by others, such as Creative Commons CC-0 license. The distinguishing characteristic of OER is that it allows others to adapt the resource freely, with no or some (CC-BY, CC-BY-SA, CC-BY-SA-NC) copyright restrictions.[11]

OER's inherent value as reusable and remixable resources is to increase access, reduce costs, and enhance educational quality across populations, distances, or social statuses (Dhanarajan & Porter, 2013). OER can also help to address concrete problems, such as the availability of print resources (e.g. the high cost or the limited number of textbooks available), inadequate library and other learning facilities, and poor teacher training (Rossini, 2010).

The reported pedagogical and economic benefits of quality OER include the following: (i) it enhances both the institution's reputation and the developers' reputation, (ii) it improves performance, (iii) it enables institutions to share best practices, (iv) it reduces resource development cost and time, (v) it extends the users' knowledge of a subject or a course, (vi) it supports students with no access to higher education, (vii) it supports developing nations and (viii) it can lead to the development of communities of practice and social networks (Dhanarajan & Porter, 2013).

## OER IN ASIA

A scan of relevant literature revealed that the adoption of OERs in Asia is more widespread in IHLs as compared to K-12 education. Generally, governments have no fixed policy on OER, with OER efforts mainly led by individual IHLs or non-profit organisations and individuals.

In the case of IHLs, the various Open Universities in Asia rely on OERs for course material and/or course development. For example, University of the Philippines Open University, and Wawasan Open University (WOU) in Malaysia use OER from various sources to form course packages or study materials (Palowski, 2010).

### Singapore

A foray into OER in Singapore is represented by the Open Source Physics in Singapore (OSP@SG) project. OSP@SG is a research project by the Educational Technology Division of the Ministry of Education, Singapore, funded by a series of National Research Fund (NRF) eduLab[12] funding initiatives since 2012. The OSP@SG project has received the 2015 UNESCO King Hamad Bin Isa Al-Khalifa

---

[11] https://www.oercommons.org/static/staticpages/documents/FAQ-OER-K-12.pdf

[12] http://edulab.moe.edu.sg/edulab-programmes/existing-projects

Prize, Singapore[13] for its Pedagogical Innovation in the Use of ICT in Teaching and Learning.

The OSP@SG project helps teachers bring real-world physics concepts in and outside schools through OER. OSP@SG is a digital library containing Java, JavaScript and Tracker resources. The programme complements real-life experiments by providing interactive digital resources that run on computers and mobile devices. OSP@SG also created a mathematical modelling function, where students' mathematical ideas can be compared with real life and simulation data. As an OER, the resources developed by OSP@SG can be freely shared or adapted by teachers and students all over the world with some restrictions such as creative commons attribution CC-BY for Java and Tracker resources and creative commons attribution, share-alike, non-commercial CC-BY-SA-NC for JavaScript resources. The OSP@SG hopes to contribute to an inclusive education and promote lifelong learning, while exploring means for sustainability through charging a small fee for commercial exploits of JavaScript resources.

Building on the strengths of the Open Source Physics[14] project based in the USA and Spain, OSP@SG also contributes source codes to OERs (simulations and video tracker) that allow teachers and students to edit and republish remixed resources under creative commons attribution, share-alike, non-commercial licenses CC-BY-SA-NC. This license gives the required permission ahead and makes it clear what others can and cannot do with these OERs in order to use them.

**ISSUES AND CHALLENGES FOR OER**

While OER has value in improving the overall access to education, there remain barriers to the widespread adoption of OER at both the government and institution level, as well as, at the individual level. Challenges include issues on intellectual property, commercialisation interests, and sustainability.

### i.  Intellectual Property Rights

Existing policies and practices at the government or institution level could be contrary to the spirit of OER. One example is the intellectual property rights issues that stem from the development of OER by a user. The OER, by definition, would require a specific copyright clause, which may differ from existing institutional policies, where resources created would belong expressly to the institution with no derivatives or sharing allowed. When OER licensed as creative commons share alike are adapted to create new resources, the resources remixed by a user have to be licensed as creative commons share alike and hence, an OER as well. However, if creative commons no rights reserved license, CC0 was used, the user is not required to license the adapted resources as OER. With OER and the increasing popularity of the creative commons licensing, institutions would need to strengthen their understanding of intellectual property rights and related issues[15].

---

[13] http://www.unesco.org/new/index.php?id=129455
[14] http://www.opensourcephysics.org
[15] Open Education Resources: An Asian Perspective. G. Dhanarajan, D. Porter. (Ed.) Published by Commonwealth of Learning and OER Asia, Vancouver, 2013

### ii. Commercialisation Interests

The model of education based on commercialisation in a particular country can also be detrimental to the widespread adoption of OER. If the higher education system is highly commercialised, this may discourage OER creation and its use in IHLs[16]. Profit-driven business entities may be reluctant to open up free public access to their materials.

### iii. Sustainability

Funding policies could present another issue to the development and sustainability of OER. Project-based funding, if used to develop OER, might lead to the resource becoming obsolete when the funding duration runs out. In that case, the sustainability of OER after the funding duration would depend on the educator community continuing to develop the OER through voluntary and autonomous activities.[17]

## CONCLUSION

In promising accessible, inclusive, massive and free offerings, both MOOCs and OER usher a new paradigm to education. The popularity of MOOCs has drawn attention from institutions, governments and private investors around the globe trying to build their brands and to enter the education and training market. The draw of OER is evident from the emergence of a ground-up community of passionate developers, educators, and users in Singapore and around the world.

While the possibilities of MOOCs and OER can be attractive, it is important to remain cognisant of the implications and limitations, which these new phenomena also bring. As emerging trends that have the potential to radically shape the educational landscape, a deeper study on MOOCs and OER in Singapore, as well as their impact on policy and legislation in specific societal and national contexts, can be illuminating.

---

[16] Ibid.
[17] Ibid.